\documentstyle[aps,prl,preprint,floats,epsfig]{revtex}

\begin{document}

\preprint{\tighten\vbox{\hbox{\hfil CLNS 02/1789}
                        \hbox{\hfil CLEO 02-9}
}}

\title{Observation of $B\to K^0_S\pi^+\pi^-$ and Evidence for $B\to K^{*\pm}\pi^\mp$}  

\author{CLEO Collaboration}
\date{December 3, 2002}

\maketitle
\tighten

\begin{abstract} 
We report on a search for charmless hadronic $B$ decays to the three-body
final states $K^0_S h^+\pi^-$, $K^+ h^-\pi^0$, $K^0_S h^+\pi^0$
($h^\pm$ denotes a charged pion or kaon), and their
charge conjugates, using 13.5 ${\rm fb}^{-1}$ of integrated
luminosity produced near $\sqrt{s}=10.6$ GeV, and collected with
the CLEO detector.  We observe the decay $B\to K^0 \pi^+\pi^-$
with a branching fraction
$(50^{+10}_{-9}$(stat.)$\pm 7$(syst.)$)\times 10^{-6}$ and
the decay $B\to K^{*+}(892)\pi^-$ with a branching fraction
$(16^{+6}_{-5}$(stat.)$\pm 2$(syst.)$)\times 10^{-6}$.
\end{abstract}
\newpage

{
\renewcommand{\thefootnote}{\fnsymbol{footnote}}

\begin{center}
E.~Eckhart,$^{1}$ K.~K.~Gan,$^{1}$ C.~Gwon,$^{1}$ T.~Hart,$^{1}$
K.~Honscheid,$^{1}$ D.~Hufnagel,$^{1}$ H.~Kagan,$^{1}$
R.~Kass,$^{1}$ T.~K.~Pedlar,$^{1}$ J.~B.~Thayer,$^{1}$
% ...
\end{center}
 
\small
\begin{center}
$^{1}${Ohio State University, Columbus, Ohio 43210}\\
% ...
\end{center}

\setcounter{footnote}{0}
}
\newpage

Recent years have seen~\cite{otherExp}
the first observations of several two-body charmless
hadronic decays of $B$ mesons, including the four $B\to K\pi$ transitions.
These two-pseudoscalar decays have received considerable attention due to
their expected role in improving our understanding of the weak interaction
and in the extraction of the complex quark couplings described by
the Cabibbo-Kobayashi-Maskawa matrix~\cite{ckm}.
The pseudoscalar-vector analogs of these
decays, $B\to K^*\pi$ and $K\rho$, provide further constraints
on the magnitudes and phases of these couplings and present additional
opportunities for observing direct $CP$ violation~\cite{theory} and
deviations from the Standard Model.
Accurate measurement and interpretation of the rates and $CP$ asymmetries
of these quasi-two-body decays requires knowledge
of the non-resonant contributions to the same final states~\cite{theoryNR}.
In this Letter, we report on a study of $B$ decays to the three-pseudoscalar
final states, $K^0_S h^+\pi^-$, $K^+ h^-\pi^0$, and $K^0_S h^+\pi^0$
($h^\pm$ denotes a charged pion or kaon), without regard for the resonant
substructure.  For each final state, we also search for two-body
channels with intermediate vector resonances, as well as for non-resonant
production.
Results for the $K^0_S h^+\pi^-$ and $K^+ h^-\pi^0$ topologies have been
presented previously by the CLEO~\cite{CLEO} and Belle~\cite{belle}
Collaborations.
The inclusion of charge conjugate states is always implied.

The data sample used in this analysis was produced in symmetric $e^+e^-$
collisions at the Cornell Electron Storage Ring (CESR)
and collected with the CLEO detector in two configurations, known as
CLEO II~\cite{cleonim} and CLEO II.V~\cite{iivnim}.
It comprises 9.12 ${\rm fb}^{-1}$ of integrated luminosity
collected on the $\Upsilon(4S)$ resonance, corresponding to $9.7\times 10^6$
$B\bar{B}$ pairs, of which $6.3\times 10^6$ were taken with CLEO II.V.
An additional 4.36 ${\rm fb}^{-1}$ collected
below the $B\bar{B}$ production threshold
is used to study non-$B\bar{B}$ backgrounds.
The response of the experimental apparatus is studied with a
GEANT-based~\cite{geant} simulation of the CLEO detector, where the
simulated events are processed in a fashion similar to data.

In CLEO II,
the momenta of charged particles are measured with a
tracking system consisting of a six-layer straw
tube chamber, a ten-layer precision
drift chamber, and a 51-layer main drift chamber, all operating
inside a 1.5 T superconducting solenoid.  The main drift chamber
also provides a measurement of specific ionization energy loss ($dE/dx$),
which is used for particle identification.
For CLEO II.V,
the six-layer straw tube chamber was replaced by a three-layer
double-sided silicon vertex detector, and the gas in the main 
drift chamber was changed from an argon-ethane to a helium-propane mixture.
Photons are detected with a 7800-crystal CsI electromagnetic calorimeter,
which is also inside the solenoid.
Proportional chambers placed at various depths within the steel return
yoke of the magnet identify muons.

Charged tracks are required to be well-measured and to satisfy 
criteria based on the track fit quality and must be consistent with
coming from the interaction point in three dimensions.
Pions and kaons are identified by $dE/dx$, and tracks
that are positively identified as electrons or muons are not allowed to form
the $B$ candidate.  We form $\pi^0$ candidates from
pairs of photons with an invariant mass within 20 MeV/$c^2$ or approximately
2.5 standard deviations ($\sigma$) of
the known $\pi^0$ mass. These are then kinematically fitted with the mass
constrained to the known $\pi^0$ mass.  We also require the $\pi^0$
momentum to be greater than 1 GeV/$c$
to reduce combinatoric background from low-momentum $\pi^0$ candidates.
$K^0_S$ candidates are selected from pairs of tracks with invariant
mass within 10 MeV/$c^2$ or $2.5\sigma$
of the known $K^0_S$ mass.  In addition, $K^0_S$ candidates are required
to originate from the beam spot and to have well-measured displaced
decay vertices.

We identify $B$ meson candidates by their invariant mass and the total
energy of their decay products.  We calculate a beam-constrained mass by
substituting the beam energy ($E_b$) for the measured $B$
candidate energy: $M \equiv\sqrt{E_b^2 - {\mathbf p}_B^2}$, where
${\mathbf p}_B$ is the $B$ candidate momentum.  Performing this substitution
improves the resolution of $M$ by one order of magnitude, to about
3 MeV/$c^2$.  We define $\Delta E\equiv E_1 + E_2 + E_3 - E_b$, where
$E_1$, $E_2$, and $E_3$ are the energies of the $B$ candidate daughters.
For final states with a $K^0_S$ and two charged tracks, the $\Delta E$
resolution is about 20 MeV for CLEO II and 15 MeV for CLEO II.V.  A $\pi^0$
in the final state degrades this resolution by approximately a factor of two.
$\Delta E$ is always calculated assuming the $h^\pm$ is a pion.
Therefore, the $\Delta E$ distribution for pions is centered at zero,
while that for kaons is shifted by at least $-40$ MeV.
We accept $B$ candidates with $M$ between 5.2 and 5.3 GeV/$c^2$ and with
$|\Delta E|$ less than 300 MeV for modes containing a $\pi^0$ and 200 MeV for
$K^0_S h^+\pi^-$.  This
region includes the signal region and a generous sideband for background
normalization.  We reject candidates that are consistent with the
exclusive $b\to c$ transitions $B\to D\pi$, where $D\to K\pi$,
and $B\to\psi K^0$, where $\psi\to\mu^+\mu^-$ and the muons are misidentified
as pions.

The main background in this analysis arises from $e^+e^-\to q\bar q$,
where $q = u,d,s,c$.  To suppress this background, we calculate the
angle $\theta_{\rm sph}$ between the sphericity axis~\cite{sphericity}
of the tracks and showers
forming the $B$ candidate and that of the remainder of the event.  Because of
their two-jet structure, continuum $q\bar q$ events peak strongly at
$|\cos\theta_{\rm sph}|=1$, while the more isotropic $B\bar{B}$
events are nearly
flat in this variable.  By requiring $|\cos\theta_{\rm sph}|<0.8$, we reject
83\% of the continuum background while retaining roughly the same fraction of
signal $B$ decays.  Additional separation of signal from $q\bar q$ background
is provided by a Fisher discriminant~\cite{Fisher} ${\cal F}$
formed from eleven
variables: the angle between the sphericity axis of the candidate and the
beam axis, the ratio of Fox-Wolfram moments $H_2/H_0$~\cite{r2}, and the
scalar sum of the visible momentum in nine $10^\circ$ angular bins around
the candidate sphericity axis.  We also make use of the angle between the
$B$ candidate momentum and the beam axis, $\theta_B$.  Angular
momentum conservation causes $B$ mesons produced through the $\Upsilon(4S)$ to
exhibit a $\sin^2\theta_B$ dependence, while candidates from continuum
are flat in $\cos\theta_B$.

Our loose selection criteria result in samples consisting primarily of
background events and containing 11893 candidates for $K^0_S h^+\pi^-$,
28589 for $K^+ h^-\pi^0$, and 9339 for $K^0_S h^+\pi^0$.
To extract signal yields, we perform unbinned maximum likelihood fits using
the observables
$M$, $\Delta E$, ${\cal F}$, $\cos\theta_B$, and the $dE/dx$ of the faster
of the two primary tracks
(typically with momentum above 1 GeV/$c$) in $K^0_S h^+\pi^-$ and
$K^+ h^-\pi^0$ and of the only primary track in $K^0_S h^+\pi^0$.
At high momentum, charged pions and kaons are statistically separated
by their $dE/dx$ and by $\Delta E$, each of which
provides discrimination at the $2.0\sigma$ level ($1.7\sigma$ for CLEO II),
and we fit for both particle hypotheses simultaneously.  Charged pions and
kaons with momentum below 1 GeV/$c$ are cleanly identified by $dE/dx$
consistency at the $3\sigma$ level.

We perform one fit for each topology,
$K^0_S h^+\pi^-$, $K^+ h^-\pi^0$, and $K^0_S h^+\pi^0$, allowing for 
six signal and background components, pion and kaon
hypotheses for $h^\pm$ for each of the
following: signal, continuum background, and background from $b\to c$ decays.
The probability for an event to be consistent with a given component is
the product of the probability distribution function (PDF) values for each of
the input variables (neglecting correlations).  The likelihood for each event
is the sum of
probabilities over the six components, with relative weights determined by
maximizing the total
likelihood of the sample, which is given by the following expression:
${\cal L} = \prod_{i=1}^{\# {\rm events}} \left[
\sum_{j=1}^{\# {\rm components}}
\left( f_j \prod_{k=1}^{\# {\rm variables}} {\cal P}_{ijk} \right) \right]$.
The ${\cal P}_{ijk}$ are the per event PDF values, and the $f_j$ are the
free parameters optimized by the fit and constrained to sum to unity.
Since the PDFs are normalized to unit integral over the fiducial region,
the $f_j$ can be interpreted as component fractions.
The parameters of the $dE/dx$ PDFs are measured from $D\to K^+\pi^-$ decays
in data. For all other variables, the signal and $b\to c$ PDFs are determined
from high-statistics Monte Carlo samples, and the continuum PDFs are determined
from data collected below the $B\bar{B}$ production threshold.  
The impact of correlations among the input variables is reduced by determining
the PDFs as a function of the event location in the Dalitz plot,
for coarse bins in the $M^2(K\pi)$-$M^2(\pi\pi)$ plane.
We use Monte Carlo simulation to estimate the
systematic error associated with neglecting any remaining correlations.

Detection efficiencies and crossfeed among the signal modes are
measured from Monte Carlo simulated events.
The statistical significance of the raw yield $N$ reported by the fit
is determined by repeating the
fit with $N$ fixed to be zero.  A significance of $n\sigma$
results in an increase in $-2\ln{\cal L}$ of $n^2$.
We calculate fit yield upper limits ($N^{UL}$) at the 90\% confidence level
by integrating the likelihood function:
$        \frac{\int^{N^{UL}}_{0} {\cal L}_{\rm max}(N) dN}
                        {\int^{\infty}_{0} {\cal L}_{\rm max}(N) dN} = 0.90$,
where ${\cal L}_{\rm max}(N)$ is the maximum likelihood at fixed $N$,
which conservatively accounts for
possible correlations among the free parameters in the fit.

Table~\ref{table:resultsIncl} lists the results of the fits to the three
topologies.  We observe a signal for $B\to K^0_S\pi^+\pi^-$
with a statistical significance of $8.1\sigma$.
Since the efficiency depends on position in the Dalitz plot,
we evaluate the efficiency in bins across the
Dalitz plot and apply a correction to each event.
The branching fractions and upper limits thus obtained are free from model
dependence.  The efficiencies given in Table~\ref{table:resultsIncl}
are averages over the efficiencies for the
observed events weighted by the probability that they are signal.
For modes with yields consistent with zero, {\it i.e.} the three $KK\pi$ modes,
the signal Dalitz plot distribution is unknown.
Therefore, we examine several models of signal distribution across the Dalitz
plot, both resonant and non-resonant, and we adopt the model with the lowest
measured efficiency to establish conservative upper limits.
The two sets of errors correspond to the statistical and systematic
uncertainties, respectively.  The largest contributions to the latter are
uncertainties in the PDF parameters and reconstruction efficiencies.
Branching fraction upper limits are increased by one standard deviation
of the relative systematic uncertainty.
We assume equal branching fractions for $\Upsilon(4S)\to B^0\bar{B}^0$
and $B^+B^-$, which is consistent with a recent measurement~\cite{f00}.

In these six-component fits, we measure yields for the three-body
final states ignoring any possible resonant substructure.  We also perform
fits for each topology that include the Dalitz plot variables as inputs
and allow for various intermediate resonances
($K^*(892)$, $K^*_0(1430)$, $\rho(770)$, and $f_0(980)$) as well as
non-resonant phase space decay.
The Dalitz plot PDFs include our knowledge of the helicity structure in
these decays.
We neglect interference among these processes and assign a systematic
error estimated from Monte Carlo simulation.
The decays $B\to K^+(X) h^-$, where $K^+(X)$ denotes $K^{*+}(892)$ or
$K^{*+}_0(1430)$,
are accessible through 
different $K^+(X)$ submodes in both
the $K^0_S h^+\pi^-$ and $K^+ h^-\pi^0$ topologies.  
To maximize our sensitivity to these
decays, we perform the Dalitz plot
fits for these two topologies simultaneously, with the branching fraction
for each $K^+(X) h^-$ decay constrained to be equal in its two $K^+(X)$
submodes, which are related by isospin.

We perform Dalitz plot fits for the three topologies with differing
combinations of intermediate resonant and non-resonant states, with up to nine
signal components.  The only channel where we observe a statistically
significant signal is $B\to K^{*+}(892)\pi^-$ with a yield of
$12.6^{+4.6}_{-3.9}$
for $K^{*+}(892)\to K^0\pi^+$ and $6.1^{+2.2}_{-1.9}$ for
$K^{*+}(892)\to K^+\pi^0$ and
a combined significance of $4.6\sigma$.  The size of the
signal is insensitive to the choice of other resonances included in the fit.
In the $K^0_S h^+\pi^-$ topology, we find the fractional yield
$N(K^{*+}(892)\pi^-)/N(K^0_S\pi^+\pi^-)$ to be $0.20^{+0.08}_{-0.07}$,
which is $7.4\sigma$ from unity.
With efficiencies of 8.1\% and 3.9\% for $K^{*+}(892)\to K^0\pi^+$ and
$K^+\pi^0$, respectively, we obtain a branching fraction of
${\cal B}(B\to K^{*+}(892)\pi^-) = (16^{+6}_{-5}\pm 2)\times 10^{-6}$.
Most theoretical predictions~\cite{theory}
for this branching fraction lie in the range (2--14)$\times 10^{-6}$.

\begin{figure}[H]
\begin{center}
\epsfig{file=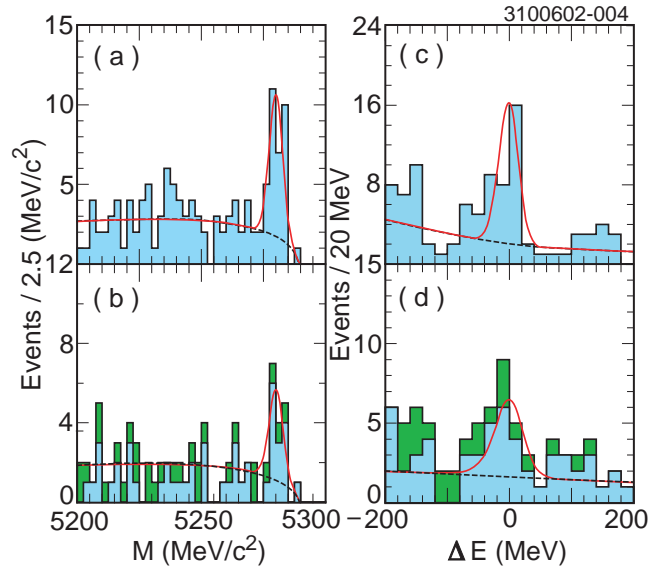,scale=0.5}
\caption{$M$ and $\Delta E$ projections for
$B\to K^0\pi^+\pi^-$ (a and c) and $B\to K^{*+}(892)\pi^-$ (b and d), which
include
the two $K^{*+}(892)$ submodes, $K^{*+}(892)\to K^0\pi^+$ (light shade) and
$K^{*+}(892)\to K^+\pi^0$ (dark shade). The background
has been suppressed with cuts on the ratios of signal and background
likelihoods computed without the displayed variable.
The dashed and solid lines show the fit
predictions for background and the sum of signal and background, respectively.}
\label{figure:projections}
\end{center}
\end{figure}

\begin{figure}[H]
\begin{center}
\epsfig{file=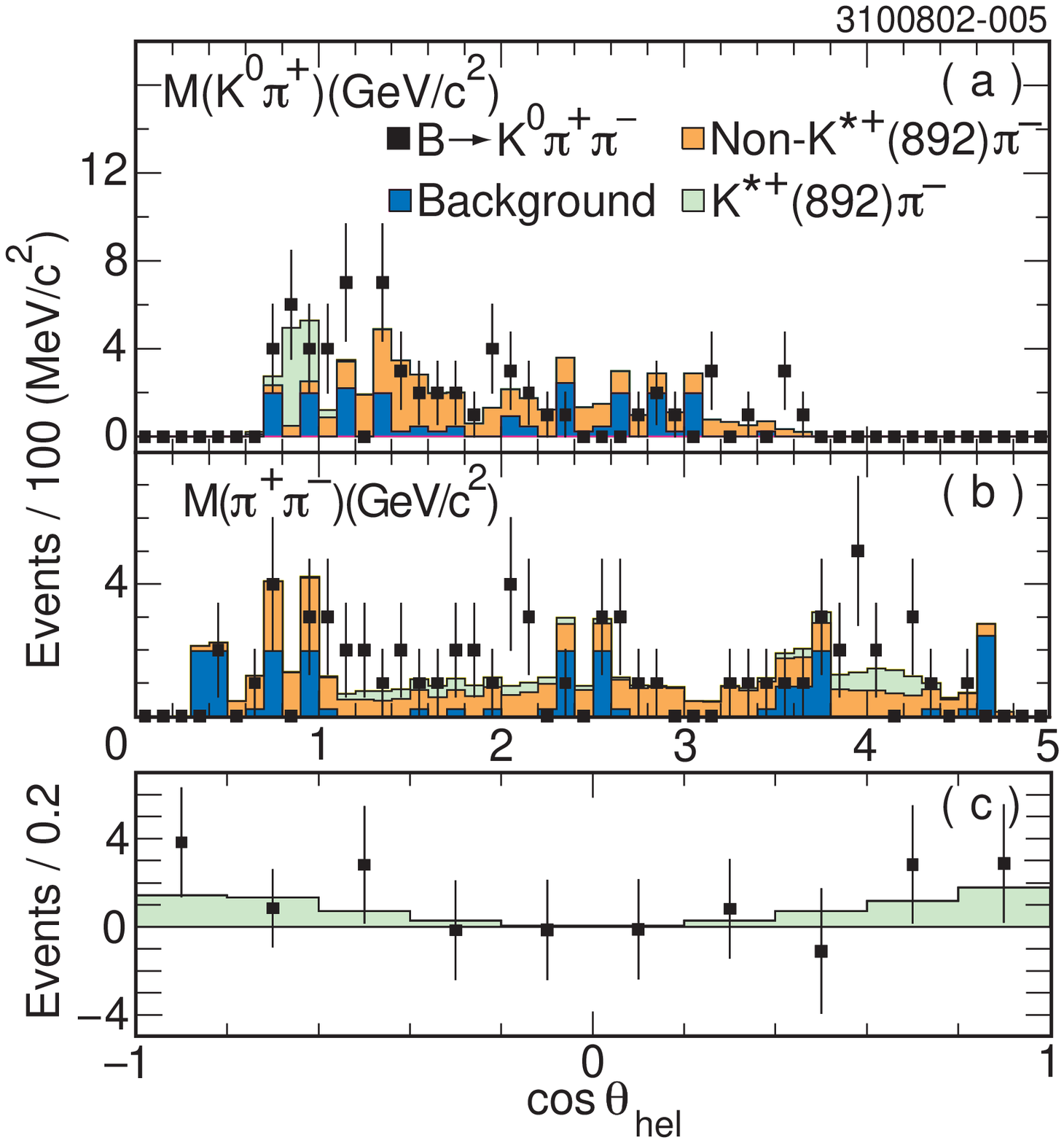,scale=0.5}
\caption{$M(K^0\pi^+)$ and $M(\pi^+\pi^-)$ projections for
$B\to K^0\pi^+\pi^-$ (a and b, respectively) and the
$\cos\theta_{\rm hel}$ projection for $B\to K^{*+}(892)(K^0\pi^+)\pi^-$ (c).
Shown in (a) and (b) are the
distribution for data (points) and the fit predictions for
background (dark shade), the sum of all signal $B$ decays except
$B\to K^{*+}(892)\pi^-$ (medium shade), and $B\to K^{*+}(892)\pi^-$ (light
shade).  Shown in (c) are the data distribution (points) in the region
$0.75< M(K^0\pi^+) < 1.05$ GeV/$c^2$ with all non-$K^{*+}(892)\pi^-$
contributions subtracted and the fit prediction for
$B\to K^{*+}(892)\pi^-$ (histogram).  The background
has been suppressed with a cut on the ratio of signal and background
likelihoods computed without the Dalitz plot variables.}
\label{figure:mkst}
\end{center}
\end{figure}

Figure~\ref{figure:projections} shows the event distributions in $M$ and
$\Delta E$ for modes where we claim an observation.  The background in these
plots has been suppressed with cuts on the ratios of signal and background
likelihoods computed without $M$ and $\Delta E$ respectively.
Overlaid are the fit projections for the signal and background
components, scaled by the efficiency of the likelihood ratio requirements
(40--50\% for $K^0\pi^+\pi^-$ and 70--80\% for $K^{*+}(892)\pi^-$).
Figure~\ref{figure:mkst}a and~\ref{figure:mkst}b show the
$M(K^0\pi^+)$ and $M(\pi^+\pi^-)$ distributions for events in
the $K^0_S h^+\pi^-$ fit satisfying a likelihood ratio requirement.
Overlaid are the fit predictions for background, $B\to K^{*+}(892)\pi^-$,
and all other signal modes combined.
We define the helicity angle for $B\to K^{*+}(892)(K^0\pi^+)\pi^-$,
$\theta_{\rm hel}$,
to be the angle between the $K^{*+}(892)$ daughter $\pi^+$ direction in the
$K^{*+}(892)$ rest frame and the $K^{*+}(892)$ direction in the $B$
rest frame.  Figure~\ref{figure:mkst}c shows the distribution of
$\cos\theta_{\rm hel}$ in the region $0.75 < M(K^0\pi^+) < 1.05$ GeV/$c^2$
after subtracting all contributions except
$B\to K^{*+}(892)\pi^-$, which are estimated from data collected below the
$B\bar{B}$ production threshold and from Monte Carlo
simulation.  The data are consistent with the $\cos^2\theta_{\rm hel}$
dependence expected for a pseudoscalar-vector $B$ decay and reflected
in the overlaid fit projection.

In summary, we have observed the three-body decay
$B\to K^0\pi^+\pi^-$ with a branching fraction
$(50^{+10}_{-9}\pm 7)\times 10^{-6}$.  A simultaneous analysis of the
Dalitz plots for this decay and for $B\to K^+\pi^-\pi^0$
reveals the presence of $B\to K^{*+}(892)\pi^-$ with
a branching fraction $(16^{+6}_{-5}\pm 2)\times 10^{-6}$, which is larger
than but consistent with most theoretical predictions.

\begin{table}[ht]
\caption{Maximum likelihood fit results for three-body decays.
Reconstruction efficiencies include all daughter branching fractions.
The errors on branching fractions ${\cal B}$ are statistical and systematic.
Upper limits are computed at the 90\% confidence level.}
\begin{center}
\begin{tabular}{lcccc}
Mode & Raw Yield $N$ & Significance & $\epsilon$ (\%) & ${\cal B}\times 10^{6}$ \\
\hline
$K^0\pi^+\pi^-$ & $60.2^{+11.5}_{-10.6}$ & $8.1\sigma$ &
	12 & $50^{+10}_{-9}\pm 7$ \\
$K^0 K^-\pi^+$ & $2.4^{+7.1}_{-2.4}$ & $0.4\sigma$ &
	8.0 & $< 21$ \\
\hline
$K^+\pi^-\pi^0$ & $43.0^{+14.5}_{-13.5}$ & $3.7\sigma$ &
	19 & $< 40$ \\
$K^+K^-\pi^0$ & $0.0^{+11.5}_{-0.0}$ & $0.0\sigma$ &
	14 & $< 19$ \\
\hline
$K^0\pi^+\pi^0$ & $20.3^{+10.1}_{-8.8}$ & $2.7\sigma$ &
	6.8 & $< 66$ \\
$K^0 K^-\pi^0$ & $0.0^{+3.7}_{-0.0}$ & $0.0\sigma$ &
	3.7 & $< 24$ \\
\end{tabular}
\label{table:resultsIncl}
\end{center}
\end{table}

\begin{acknowledgements}
We thank Matthias Neubert and Alex Kagan for useful discussions.
We gratefully acknowledge the effort of the CESR staff in providing us with
excellent luminosity and running conditions.
This work was supported by 
the National Science Foundation,
the U.S. Department of Energy,
the Research Corporation,
and the Texas Advanced Research Program.
\end{acknowledgements}

\renewcommand{\baselinestretch}{1}

\end{document}